\documentclass[showpacs,aps,amssymb,floatfix,prd,preprintnumbers]{revtex4-2}
\usepackage{epstopdf}
\usepackage{capt-of}
\usepackage{graphicx}  
\usepackage{dcolumn}   
\usepackage{bm}
\usepackage{amsmath}
\usepackage[font=scriptsize]{caption}
\usepackage[colorlinks]{hyperref}
\begin{document}
\input epsf.tex
\title{\bf Bouncing cosmological model with general relativistic hydrodynamics in extended gravity}

\author{A.Y.Shaikh \footnote{Department of Mathematics, Indira Gandhi Mahavidyalaya, Ralegaon 445402,India,E-mail:shaikh\_ 2324ay@yahoo.com}, Sankarsan Tarai \footnote{Centre of High Energy and Condensed Matter Physics, Department of Physicss, Utkal University, Bhubaneswar 751004, India, E-mail:tsankarsan87@gmail.com}, S.K. Tripathy \footnote{Department of Physics, Indira Gandhi Institute of Technology, Sarang, Dhenkanal, Odisha-759146, India, tripathy\_sunil@rediffmail.com} B. Mishra \footnote{Department of Mathematics, Birla Institute of Technology and Science-Pilani, Hyderabad Campus, Hyderabad-500078, India, email: biv@hyderabad.bits-pilani.ac.in}, }

\affiliation{ }

\begin{abstract}
\textbf{Abstract:}
In this paper, in an extended theory of gravity, we have presented bouncing cosmological model at the backdrop of an  isotropic, homogeneous space-time, in presence of general relativistic hydrodynamics (GRH). The scale factor has been chosen in such a manner that with appropriate normalization, the quintom bouncing scenario can be assessed. Accordingly, the bounce occurs at $t=0$ and the corresponding Hubble parameter vanishes at the bounce epoch. The equation of state (EoS) parameter and the energy conditions of the model have been analysed. The violation of strong energy condition further supports the behaviour of extended gravity. As the bouncing cosmology suffers with instability, this model also shows the similar behaviour.

\end{abstract}

\keywords{}
\maketitle
\textbf{Keywords}: Extended gravity, Bouncing cosmology, General relativistic hydrodynamics, Stability analysis.

\section{Introduction} \label{SectionI}
The fundamental issues in cosmology such as horizon, baryon asymmetry, flatness problem, initial singularity and most recent the dark energy and dark matter makes the standard cosmological model in a fix in spite of its several successes. To resolve the issue related to initial singularity, the bouncing cosmological models are being designed and analysed in recent years. According to this, the universe prevailed before the big bang and then near the non-vanishing minimum radius, it experienced the accelerated expansion phase. The change from the cosmic expansion regime to the current accelerating expansion phase is known as big bounce. The idea behind this theory is that the expansion phase begins immediately after the contraction phase resulted in a bounce. This may be helpful to provide some mechanism to resolve some of the issues faced by standard cosmological model without the inflationary scenario. But the cosmological model be well accepted if it is capable of being resolve the issues as answer by the inflationary mechanism. It is important to mention here that most inflationary scenarios can give the scale-invariant spectrum of the cosmological oscillations\cite{Contreras17}. The standard cosmological model may find the solutions to the problems occurred during the contraction before the occurrence of the bounce. At the same time, the horizon problem, can be resolved if the separated regions of the present Universe would have been in the causal connection during previous contraction phase.\\

The cosmological observations suggest that our universe is expanding and the expansion is accelerating at least at the late phase of  evolution \cite{Riess98,Perlmutter99,Bennett03}. Theoretically, these observations can be dealt with the postulation that certain exotic matter with negative pressure dominates the present epoch of the universe. So, the cosmological acceleration can be introduced via fluid with negative pressure. This leads to the unknown form of energy, known as dark energy that accounts for about $68\%$  of total mass-energy budget of the universe. Several candidates for dark energy have been proposed such as the cosmological constant\cite{Peebles03}, quintessence\cite{Ratra88,Caldwell98,Sami03}, k-essence\cite{Picon01,Chiba02,Scherrer04}, tachyon\cite{Sen02a,Sen02b,Gibbons02}, phantom\cite{Caldwell02,Elizalde04,Cline04}, holographic dark energy\cite{Horava00}, extra dimensions\cite{Rogatko04} and so on. Among these the cosmological constant is the simplest form of the dark energy candidate. Theoretically, another approach can be used to address this late time acceleration issue i.e. by the geometrical extension of general relativity. The extended theories of gravity can be considered as a new paradigm to address the shortcomings encountered by General Relativity to address the late time cosmic acceleration issue \cite{Capozziello11}. For details on the the cosmological and astrophysical applications of extended theories of gravity, one can refer \cite{Capozziello08}. Some interesting theoretical dark energy models can be seen in \cite{Bamba12}, where the rip cosmology, $\Lambda$CDM, quintessence and phantom cosmology have been discussed. A systematic review on the development of modified gravity on late time acceleration, inflation and bouncing cosmology is available in \cite{Nojiri17}. One of such extension with minimal matter-geometry coupling is the $f(R,T)$ gravity, where $R$ and $T$ respectively denote the Ricci scalar end trace of energy momentum tensor. This has been successful to some extent to address this issue of accelerating universe. Several studies are made in $f(R,T)$ gravity to address the cosmological and astrophysical issues that includes the late time acceleration. \\

We shall discuss some of the bouncing models available in the literature. Solomons et al. \cite{Solomons06} have shown the bouncing behaviour in an anisotropic universe. In Rastall's gravity, Silva et al. \cite{Silva13} have given the bouncing solutions in a barotropic fluid. Sadatian \cite{Sadatian14} in a Chaplygin gas dark energy model studied the bouncing universe and rip singularity. Brevik et al. \cite{Brevik14} have obtained the bouncing universe in an inhomogeneous dark fluid coupled with dark matter. Singh et al. \cite{Singh16} studied the bouncing universe by considering the matter field in the form of viscous fluid. To provide a description of the very early universe, Bradenberger and Peter \cite{Brandenberger17} have reviewed the status of bouncing cosmologies as alternatives to cosmological inflation. Minas et al. \cite{Minas19} in the framework of general modified gravities investigated the matter bounce scenario. Bouncing models in modified theories of gravity are also discussed in the literature. Cai et al. \cite{Cai11} have studied the matter bounce scenario in $f(T)$ gravity and investigated the scalar and tensor modes of cosmological perturbations. Bamba et al.\cite{Bamba14} derived the bouncing scenario for the modified gravities such as $f(R)$ gravity and $f(R)$ bi-gravity whereas,  Bamba et al. \cite{Bamba16} have investigated the  bounce inflation model in the framework of $f(T)$ gravity. Mishra et al. \cite{Mishra19} have obtained the bounce solution in $f(R,T)$ gravity at the backdrop of an anisotropic space-time whereas Shaikh and Mishra \cite{Shaikh20} have shown the bouncing behaviour in GRH model.
 Bouncing cosmology in non-isometric theories has been discussed  in \cite{Bajardi20}. Tripathy et al. \cite{Tripathy21} have studied the bouncing cosmology and have shown the instability of the model near the bounce. Agrawal et al. \cite{Agrawal22a} have shown the matter bounce scenarion in $f(R,T)$ gravity and its reconstruction as dark energy model \cite{Agrawal22b}. In the nonmetricity based gravity Agrawal et al. have shown the matter bounce scenario with an assumed form of the scale factor \cite{Agrawal21}.\\

We shall study the dynamical and evolutionary behaviour of the Universe. The field equations are highly non-linear and in order to solve it, certain mathematical techniques to be used. The general relativistic hydrodynamics (GRH) and magneto-hydrodynamics equations (MHD) coupled to Einstein's equations would be technically simplify the process to obtain the solution. Here we are motivated to investigate in extended theory of gravity, the bouncing scenario in presence of GRH with an isotropic space-time. The organization of the paper is as follows: Section \label{SectionII} deals with the basic formalism of extended gravity along with the brief introduction of the GRH and the field equations. In Section \label{SectionIII}, the bouncing cosmology has been presented and the dynamical parameters are investigated in Section \label{SectionIV}. The discussions  on the geometrical diagnostics and stability analysis are performed in Section \label{SectionV} along with conclusion of the work done.
\section{Basic Formalism and Field Equations} \label{SectionII}
The action for the geometrically extended $f(R,T)$ gravity can be expressed as \cite{Harko11},
\begin{equation}\label{eq:1}
S=\int \left[\frac{ \sqrt{-g}}{2}[R+\beta f(T)]+ \sqrt{-g}\mathcal{L}_m\right] d^{4}x,
\end{equation}
where $f(T)$   is an arbitrary function of the trace  $T$ of the energy momentum tensor, $\beta$ be the coupling constant  and $\mathcal{L}_m$  is the matter Lagrangian. Varying the action $S$ with respect to $g_{\mu \nu}$, the field equations of $f(R,T)$ gravity can be expressed as, 
\begin{equation}\label{eq:2}
R_{\mu \nu}-\frac{1}{2}R g_{\mu \nu}=T_{\mu \nu}-\beta f_{T}(T)T_{\mu \nu}+\frac{1}{2}\beta f(T)g_{\mu \nu}-\beta f_{T}(T)\Theta_{\mu \nu},
\end{equation}
when $\beta=0$, eqn. \eqref{eq:2} reduces to the field equations of general relativity. The energy momentum tensor can be expressed as. 
\begin{equation}\label{eq:3}
T_{\mu \nu}= -\frac{2}{\sqrt{-g}}\frac{\delta(\sqrt{-g}\mathcal{L}_{m})}{\delta g^{\mu \nu}}
\end{equation} 			
and   	
\begin{equation}\label{eq:4}
\Theta_{\mu \nu} \equiv g^{\alpha \beta}\frac{\delta T_{\alpha \beta}}{\delta g^{\mu \nu}}
\end{equation}								

The tools of astrophysics like hydrodynamics, magneto-hydrodynamics, radiation transport \cite{Font98} and nuclear astrophysics would be required by the observational data that involves the GRH phenomena. For an informative overview of relativistic hydrodynamics, one can refer \cite{Taub78}. Several techniques are availble in the literature to solve the field equations of GRH such as, (i)generalization of Roe's approximate Riemann solver numerical method \cite{Eulderink95}, (ii) special relativistic Riemann solvers \cite{Pons98}, (iii) fully self-consistent relativistic hydrodynamics code \cite{Shibata99}. Then, formulations of the equations of GRH and MHD, along with methods for their numerical solution was reviewed extensively \cite{Font07}. Subsequently a comprehensive overview of numerical hydrodynamics and magneto hydrodynamics was presented in the context of general relativity\cite{Font08}.

We consider the stress-energy tensor in the form of perfect fluid as,

\begin{equation}\label{eq:5}
T_{\mu \nu}=\rho h u_{\mu} u_{\nu}-pg_{\mu \nu}.
\end{equation}
The relativistic enthalpy $h$ can be expressed as, $h=(1+\epsilon)+\frac{p}{\rho_0}$, where $p$, $\rho_0$ and $\epsilon$ pressure, rest mass density and specific internal energy of the fluid. To note, the rest mass energy density $\rho_0$ is different from the energy density $\rho$. A constitutive relation of the form $p=p(\rho,\epsilon)$ and the ideal fluid equation of state parameter, $p=\rho_0 \epsilon(\Gamma-1)$, $\Gamma$ be the adiabatic index has been considered. Now, we rewrite the relativistic specific enthalpy as,

\begin{equation} \label{eq:6}
h=1+\Gamma \epsilon. 
\end{equation}

We wish to mention here that $\epsilon$ is temperature-dependent, which can be confirmed on causal thermodynamics \cite{Roy96} or relativistic hydrodynamics \cite{Luciano13}. However, in this paper our motivation is to study the bouncing scenario of the Universe in presence of GRH, therefore we prefer to choose a fixed value for $\epsilon$. If we consider the matter Lagrangian as, $L_{m}=-p$, eqn. \eqref{eq:4} becomes, $\Theta_{\mu \nu}=-pg_{\mu \nu}-2T_{\mu \nu}$ and the field equations \eqref{eq:2} reduce to,
\begin{eqnarray}\label{eq:7}
R_{\mu \nu}-\frac{1}{2}R g_{\mu \nu}&=& T_{\mu \nu}+\beta f_{T}(T) T_{\mu \nu}+[f_{T}(T)p+\frac{1}{2}f(T)]\beta g_{\mu \nu} \nonumber \\
&=& [1+\beta f_{T}(T)](T_{\mu \nu}+T^{int}_{\mu \nu}) 
\end{eqnarray}  	
where, 
\begin{equation}\label{eq:8}
T^{int}_{\mu \nu}=\frac{1}{1+\beta f_{T}(T)}\left[ f_{T}(T)p+\frac{1}{2}f(T)\right]\beta g_{\mu \nu} 
\end{equation}	
It is note here eqn. \eqref{eq:8} vanishes for $\beta\rightarrow 0$. We consider the functional as, $\frac{1}{2}f(T)=T$, and subsequently obtained, $T^{int}_{\mu \nu}=\frac{1}{1+\beta}\left[ 2p+T\right]\beta g_{\mu \nu}$. Harko et al. \cite{Harko11} in their seminal work proposed three forms for the function $f(R,T)$ as (i) $f(R,T)=R+2f(T)$, (ii) $f(R,T)=f_1(R)+f_2(T)$, (iii) $f(R,T)=f_1(R)+f_2(R)f_3T)$. Here, we consider the functional form of $f(R,T)$ in such a manner that the filed equations of GR can be obtained under suitable substitution of the model parameters. One of the popular choice is; $f(R,T)=R+2\beta T$ \cite{Shamir15,Mishra18}, $\beta$ being the coupling constant, such that $f(T)=2T$. The matter part of the extended theory of gravity has been discussed above and to set the filed equations of $f(R,T)$ gravity, the geometric part needs to be addressed. So here we consider an isotropic and homogeneous FLRW space-time as, 

\begin{equation}\label{eq:9}
ds^{2}=dt^{2}-a^{2}(t)(dx^{2}+dy^{2}+dz^{2}), 
\end{equation} 			

where $a(t)$ be the scale factor. The $f(R,T)$ gravity field equations \eqref{eq:7} with GRH in the form of perfect fluid can be obtained as,

\begin{eqnarray}
2\dot{H}+3H^{2}&=&-\alpha p+\beta \rho h \label{eq:10}\\
3H^{2}&=&\alpha \rho h-\beta p,\label{eq:11}
\end{eqnarray} 

where $\alpha=(1+3\beta)$ be the redefined coupling constant and an over dot denotes the derivative with respect to cosmic time. We have expressed the field equations in Hubble term and in order to obtain the expressions for the dynamical parameters, we need to assume a form for the Hubble parameter. In recent years bouncing cosmology, an alternative to the inflationary paradigm \cite{Morris88} has been studied extensively to resolve the singularity issue, therefore, we are motivated here to study the model in the matter bounce scenario. 

\section{Bouncing Cosmological Model}\label{SectionIII}
We consider here the bouncing scale factor as $a(t)=\sqrt{1+\eta^{2}t^{2}}$ with $\eta>0$. This scale factor is the temporal analogue of the toy model of the traversable wormhole\cite{Morris88} and with proper renormalization of the scale factor parameter $\eta$, the phenomenological quintom bouncing model can be obtained \cite{Cai07}. The Hubble parameter can be expressed as 
\begin{equation}\label{eq:12}
H=\frac{\dot{a}}{a}=\frac{\eta^{2}t}{1+\eta^{2}t^{2}}
\end{equation} 

\begin{figure}[h!]
\includegraphics[width=60mm]{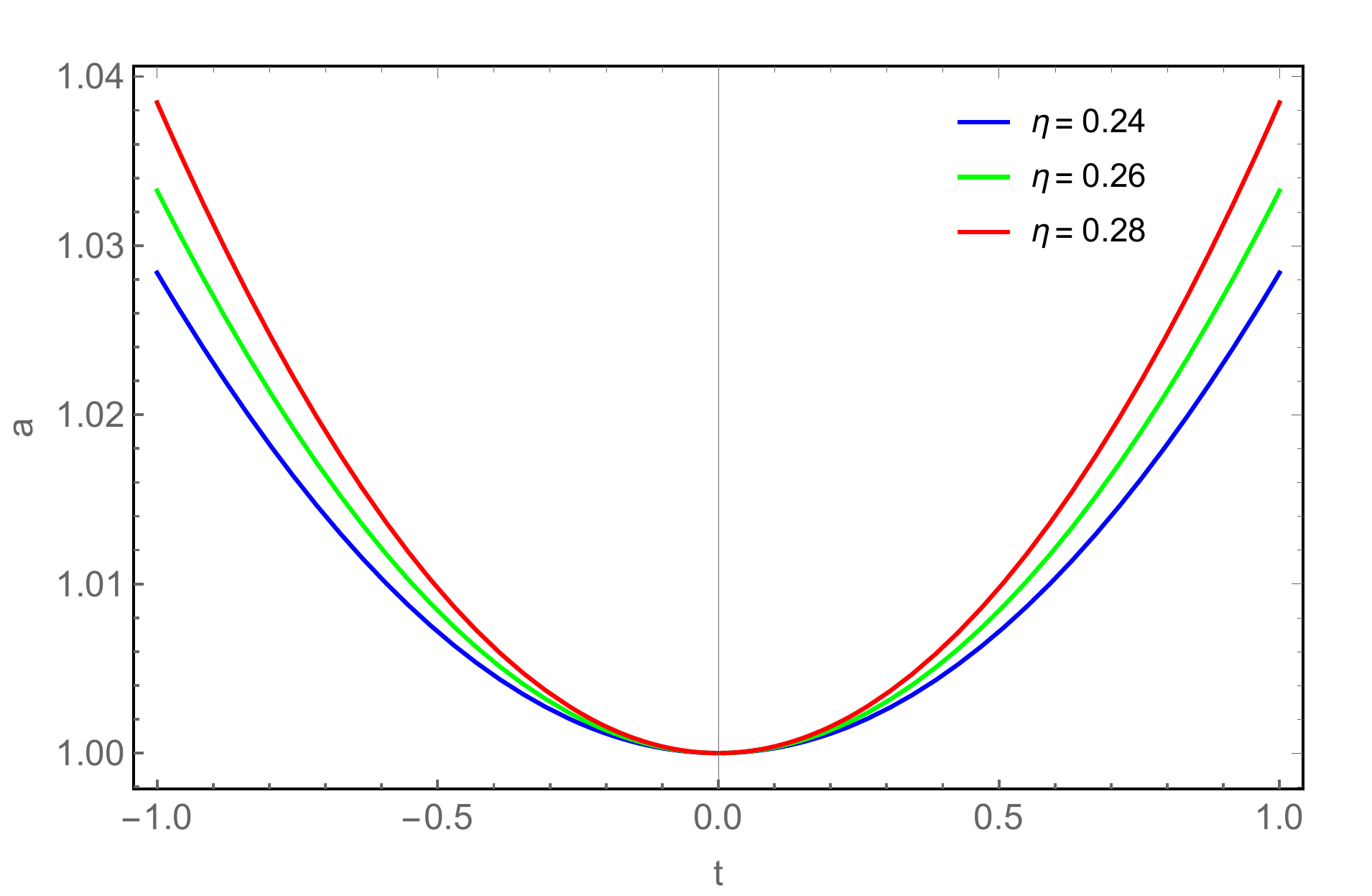}
\includegraphics[width=60mm]{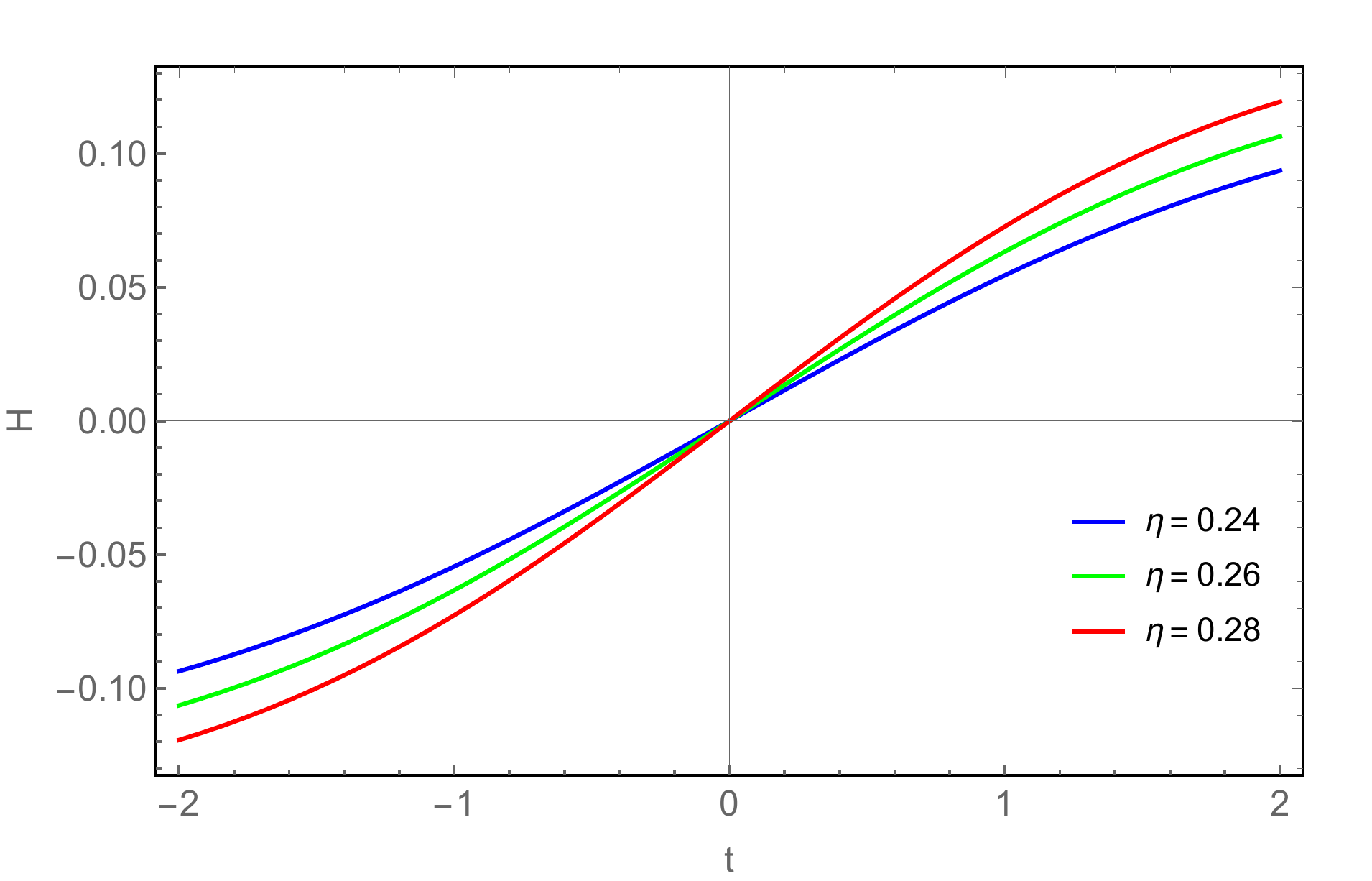}
\caption{Graphical representation of scale factor, $a$ (left panel) and Hubble parameter $H$ (right panel) in cosmic time for the representative values of $\eta$.} \label{Fig.1}
\end{figure}

We have presented the graphical behaviour of the scale factor and Hubble parameter in Fig. \ref{Fig.1} to verify the occurrence of bounce during the evolution of the universe. 
According to bouncing cosmology, the model should initially undergoes a phase of collapse, attains its minimum value at the bouncing point and then expands subsequently. In Fig. \ref{Fig.1} (left panel), the scale factor shows a symmetrical behaviour. The slope of the scale factor depends upon the parameter $\eta$. For attaining the bounce in the FLRW model space-time, $a(t)$ is decreasing i.e. $(\dot{a}(t)<0)$  during the negative frame of cosmic time (contracting universe)  and then in the expanding phase, the scale factor is increasing i.e. $(\dot{a}(t)>0)$. The bounce has been noticed at $t=0$ for positive $\eta$. At $t=0$, the scale factor curve shows the symmetric behaviour and attains a non-zero minimum value $a(t)\cong 1$ at the bounce. In a nutshell, more is the value of $\eta$, more is the curvature. In Fig. \ref{Fig.1} (right panel), we have plotted the evolution of Hubble parameter for three different values of $\eta$. As we are keeping in our mind to construct a bouncing model, we may draw the cosmic time from negative domain to positive domain and consequently the evolution of Hubble parameter goes linearly from negative to positive time domain through $t=0$ at bounce. In the negative time zone, the Hubble parameter remains negative ($H<0$)in the interval $-2<t<0$, and in the positive time zone, it is positive ($H>0$)  in the interval $0<t<2$. At the bouncing point the Hubble parameter vanishes irrespective of the value of $\eta>0$.  The deceleration parameter explains the nature of expansion of the model and can be obtained with the second derivative of the scale factor. For this bouncing sale factor, the deceleration parameter can be obtained as, 

\begin{equation}\label{eq:13}
q=-1-\frac{\dot{H}}{H^{2}}=-\frac{1}{\eta^{2}t^{2}}
\end{equation}	

Bolotin et al.\cite{Bolotin15} classified the cosmological models on the basis of time dependence on Hubble parameter and deceleration parameter. 
\begin{itemize}
\item For $H>0$, the model expands and the expansion is accelerating for $q<0$ or decelerating for $q>0$ and contracting behaviour for $H<0$. 
\item For $H>0$ and $q=0$, the model expands, and zero deceleration or constant expansion and contracts for $H>0$.
\item For $H=0$ and $q=0$, the model remains static.
\end{itemize}

\begin{figure}[h!]
\includegraphics[width=90mm]{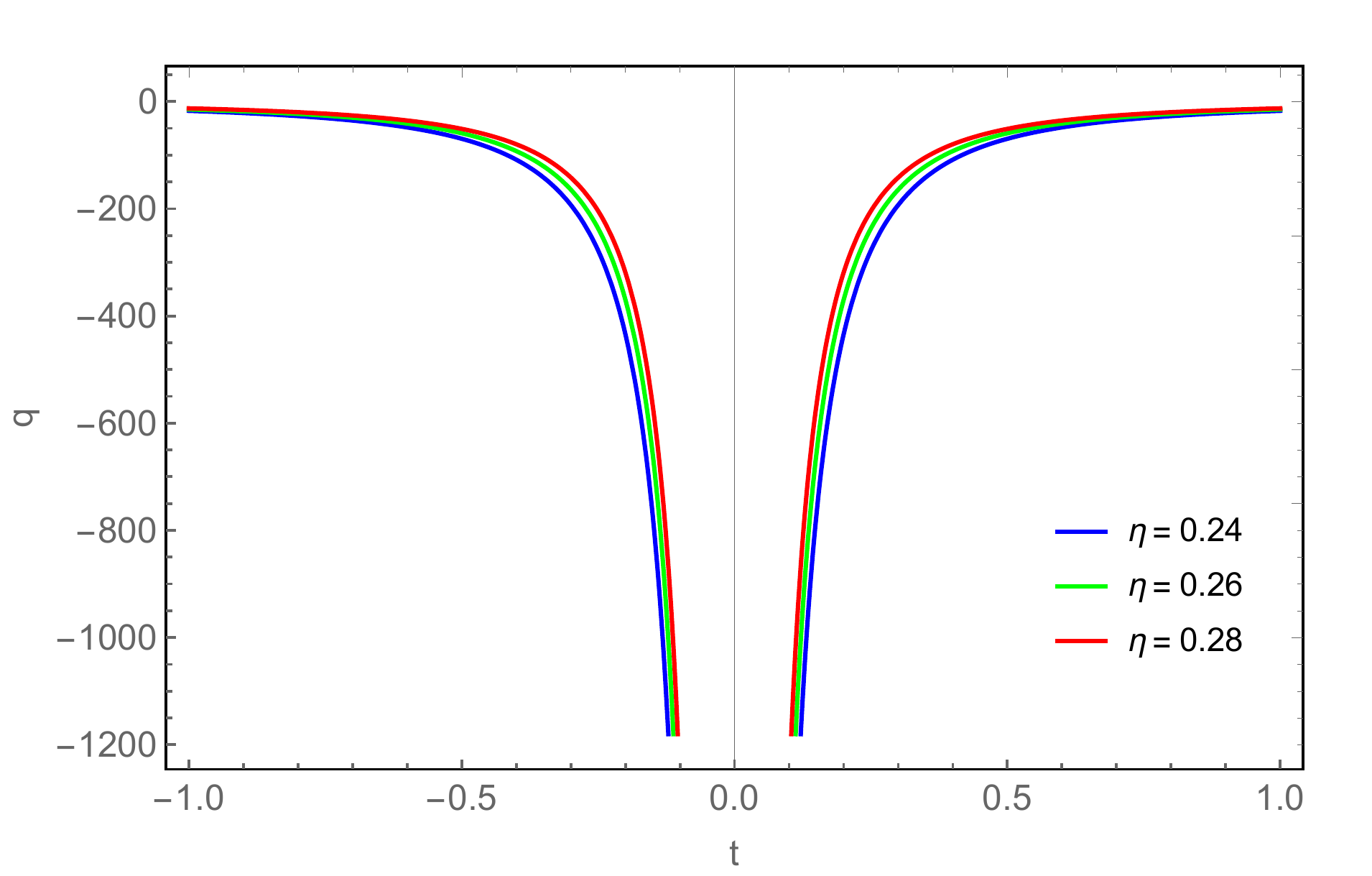}
\caption{Graphical representation of deceleration parameter in cosmic time for the representative values of $\eta$.}\label{Fig.2}
\end{figure}	 

The deceleration parameter shows symmetry behaviour as in Fig. \ref{Fig.2}, at the bouncing point, $t=0$. From Fig. \ref{Fig.2}, it is observed that for both the contracting and expanding universes, the deceleration parameter $q$  is negative, and after some finite time it tends to a constant value $–1$. In the negative time zone (contacting universe), the deceleration parameter tends to the large negative values at the bouncing point even after evolving from $q=-1$. In the positive frame of the cosmic time (expanding universe), the deceleration parameter tends to $q=-1$ at late times.\\

\section{Analysis of Dynamical Parameters of the Model}\label{SectionIV}
The viability of any cosmological model based on the appropriate behaviour of the dynamical parameters and its behaviour must be converging towards agreement. The disagreement, if any, reflect the systematic errors that comes either from the different methods of analysis or from the cosmological observations. It may also come from the representative values chosen for the model and free parameters involved in the expression.  The late time cosmic acceleration phenomena requires the pressure and equation of state (EoS) parameter to be negative at least at the present and future time. Hence, we shall  analyse here the dynamical parameters of the model in the context of bouncing behaviour. Substituting the expression of the Hubble parameter \eqref{eq:12} in eqns. \eqref{eq:10} and \eqref{eq:11} and with an algebraic manipulation, we can obtain the mater pressure and energy density as,
\begin{eqnarray}
p&=&\frac{\eta^{2}[-3(1+2\beta)\eta^{2}t^{2}+2(1+3\beta) (1-\eta^{2}t^{2})]}{(1+8\beta^{2}+6\beta)(1+\eta^{2}t^{2})^{2}} \label{eq:14} \\ 
\rho&=&\frac{\eta^{2}[3(1+2\beta)\eta^{2}t^{2}-3\beta (1-\eta^{2}t^{2})]}{(1+\Gamma \epsilon)(1+8\beta^{2}+6\beta)(1+\eta^{2}t^{2})^{2}} \label{eq:15}
\end{eqnarray}		
The EoS parameter governs the gravitational properties of dark energy model and its evolution profile. In the cosmological models, either it appears as a time dependent function or a constant. Cosmological observations \cite{Hinshaw13,Ade14} have suggested ranges for the EoS parameter as,

\begin{eqnarray*}
Planck+ WP +Union 2.1:~~~~~-0.92&\leq & \omega \leq -1.26,\\
Planck+ WP +BAO: ~~~~~~~-0.89&\leq & \omega \leq -1.38,\\
WMAP+eCMB+BAO+H0:~~~~~~~~-0.983&\leq &\omega \leq -1.162.
\end{eqnarray*}

We can derive the EoS parameter $\omega=\frac{p}{\rho}$ as,
\begin{equation}
\omega=(1+\Gamma \epsilon)\left[\frac{-3(1+2\beta)\eta^{2}t^{2}+2(1+3\beta) (1-\eta^{2}t^{2})}{3(1+2\beta)\eta^{2}t^{2}-3\beta (1-\eta^{2}t^{2})}\right] \label{eq:16}
\end{equation} 
					
\begin{figure}[h!]
\includegraphics[width=60mm]{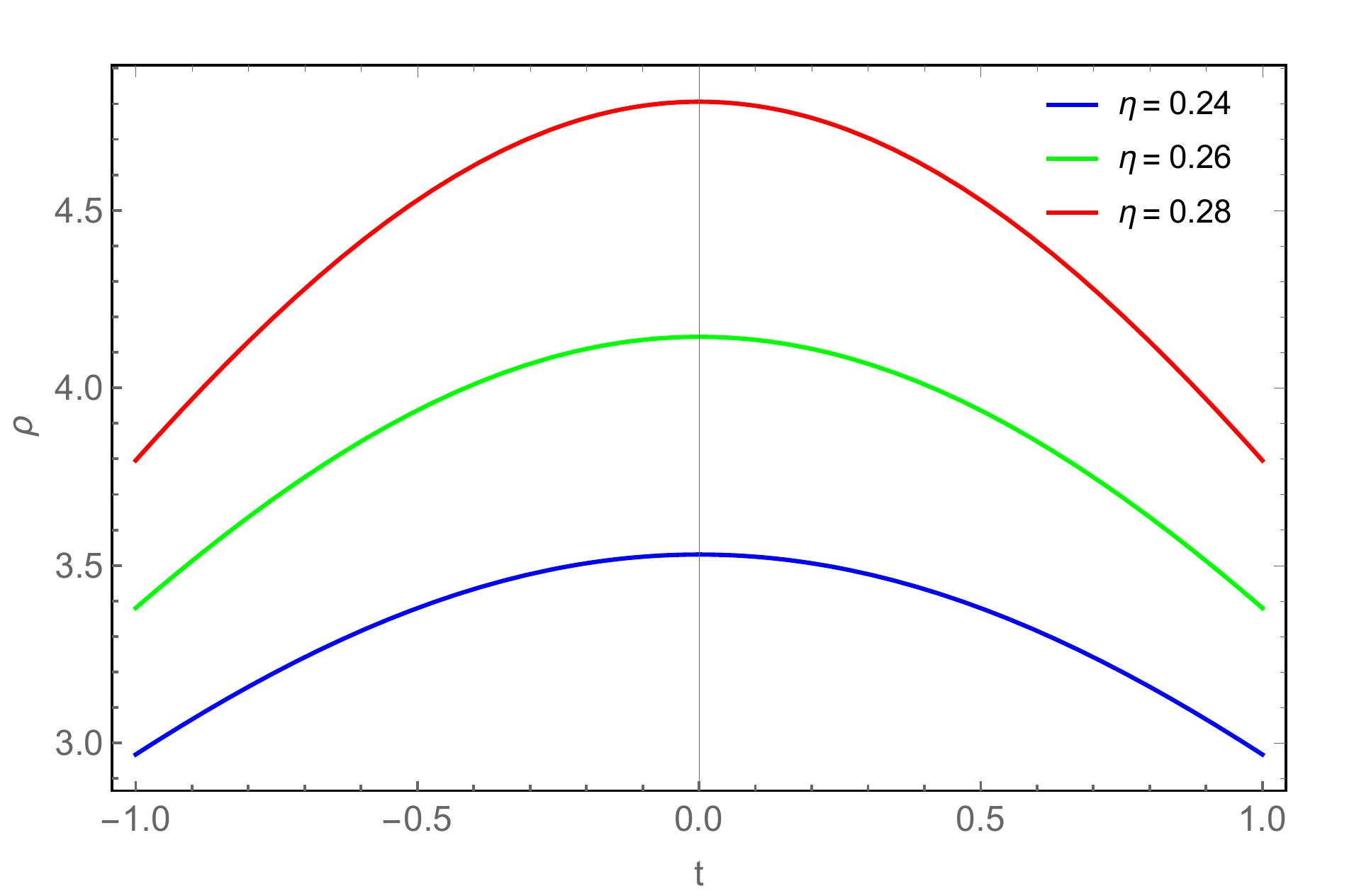}
\includegraphics[width=60mm]{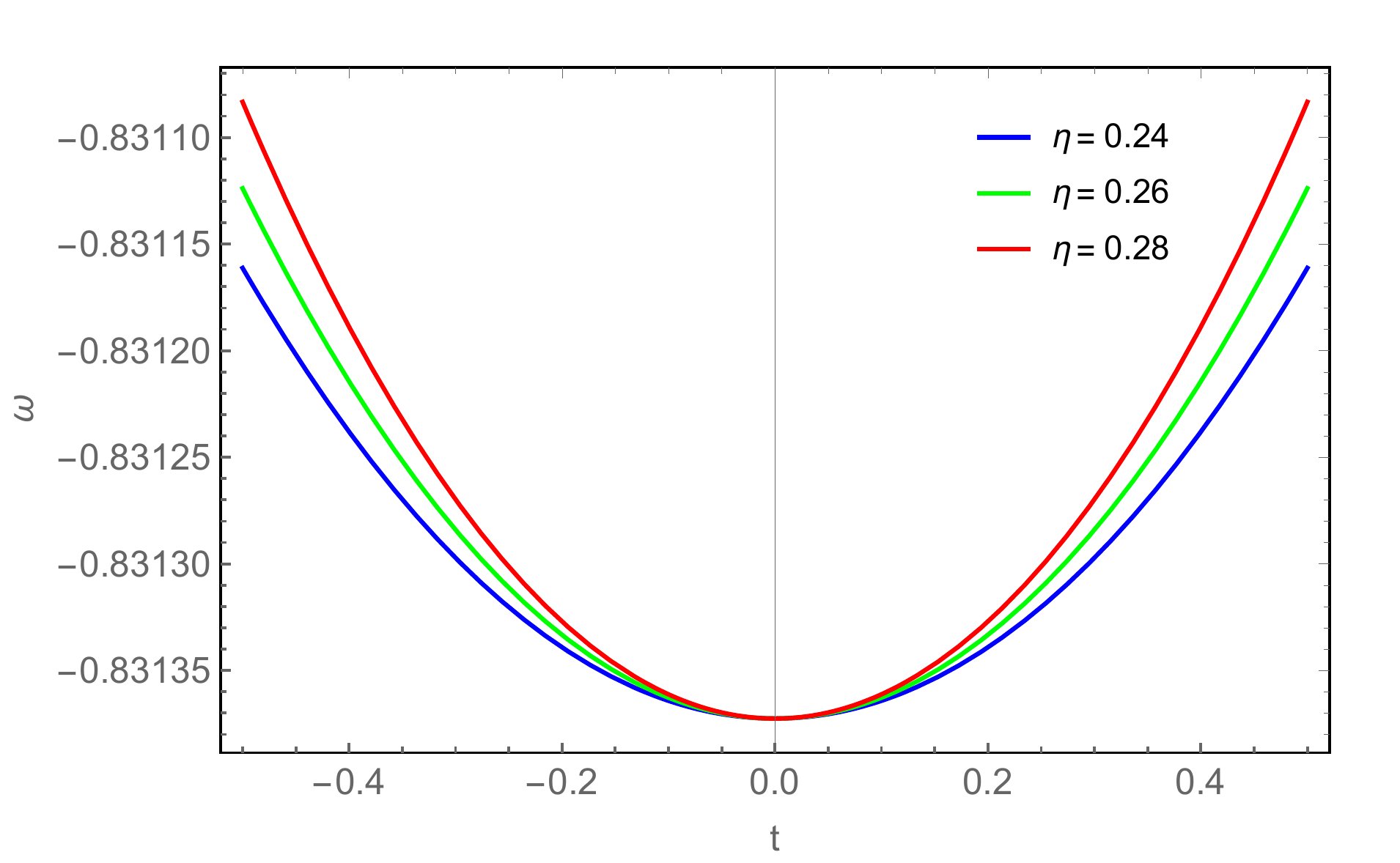}
\caption{Graphical representation of energy density (left panel) and EoS parameter (right panel) in cosmic time for the representative values of $\eta$, $\beta=-0.51$, $\Gamma=2$, $\epsilon=0.1$}\label{Fig.3}
\end{figure}
	 
From eqns. \eqref{eq:14}-\eqref{eq:15}, it is observed that the pressure and energy density of the GRH model depend on the value of the scale factor parameter $\eta$ and model parameter $\beta$. For the positive value of the model parameter $\beta$, the denominators of eqns. \eqref{eq:14}-\eqref{eq:15} are always positive. The positivity or the negativity of these quantities depends only on the respective numerators. In order to satisfy certain energy conditions, the energy density should remain positive throughout the cosmic evolution. For  $\eta=0.24$, $\beta=-0.51,$ $\Gamma=2,$ $\epsilon=0.1$ the energy density remains positive throughout the cosmic evolution both in the positive and negative domains of cosmic time. The same can be observed in Fig. \ref{Fig.3} (left panel). The evolution behaviour of EoS parameter, a time dependent function, has been shown in Fig. \ref{Fig.3} (right panel). It is observed that for $\beta \rightarrow 0$, the corresponding GRH model reproduces the results of general relativity i.e. $\omega_{\beta \rightarrow0}\approx (1+\Gamma \epsilon)\left( -1+\frac{2 (1-\eta^{2}t^{2})}{3\eta^{2}t^{2}}\right)$ . From Fig. \ref{Fig.3} (right panel), it is observed that, the EoS parameter shows symmetrical behaviour about the bouncing time. At the bounce $(t=0)$, the EoS parameter attains the value $-0.838$, which shows the quintessence behaviour of the dark energy fluid \cite{Ratra88}. The constraints of EoS parameter of the derived GRH model coincide with the WMAP nine years observational data\cite{Hinshaw13} and the Planck collaboration results \cite{Ade14}, which is specified above before eqn. \eqref{eq:16}. 

\begin{figure}[h!]
\centering
\includegraphics[width=90mm]{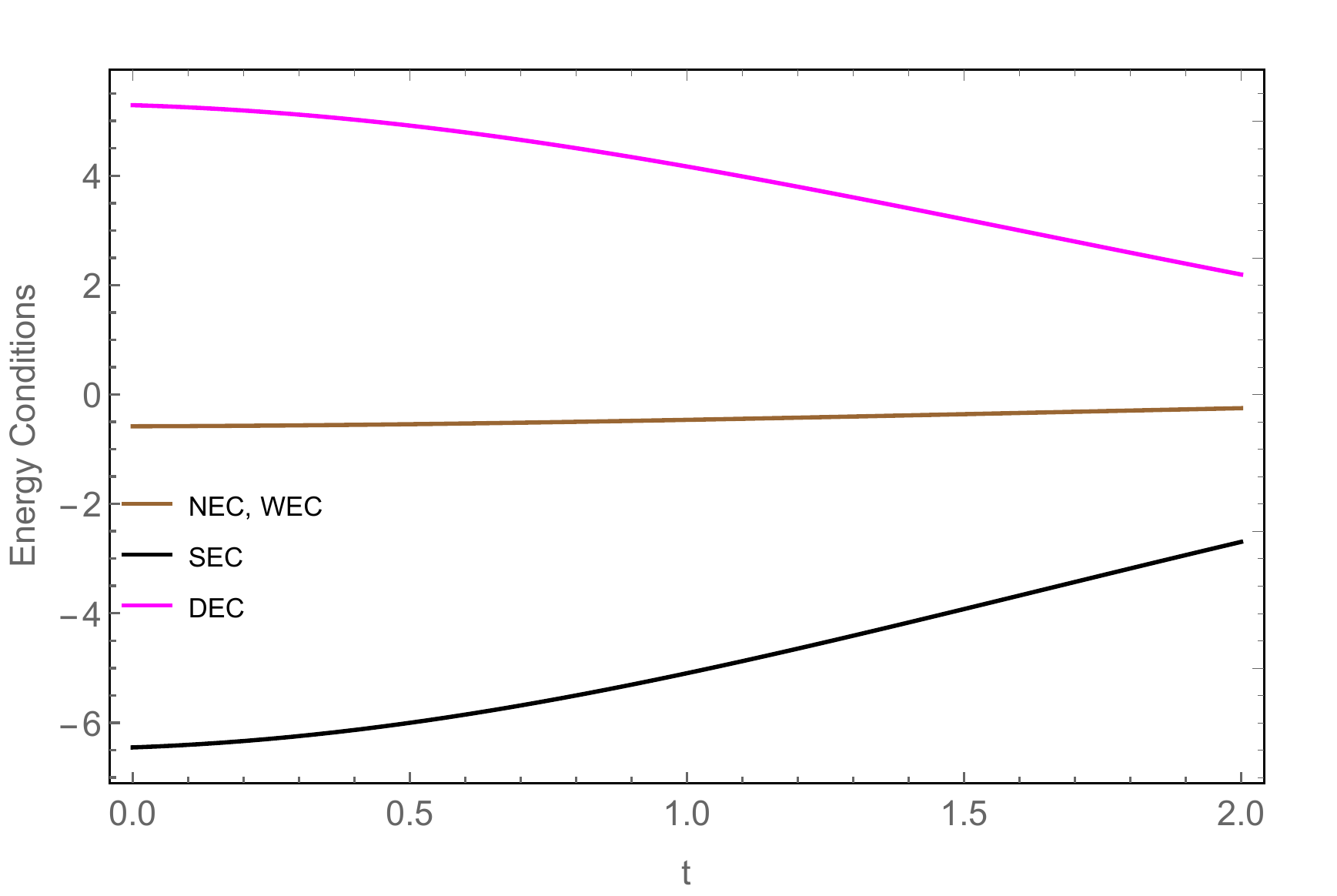}
\caption{Plot of energy conditions in cosmic time for $\eta=0.24$, $\beta=-0.51,$ $\Gamma=2,$ $\epsilon=0.1$} \label{Fig.4}
\end{figure}

Another prescription for the bouncing model is the violation of null energy condition and strong energy condition at the bounce. In general relativity, energy conditions are used extensively to examine the singularity problem of the space-time.  Essentially, the energy conditions are described by the behaviour of space like, time like or light like curves \cite{Hawking73,Wald84,Visser95}. We have considered here the space like and time like curves. The energy conditions can be defined in extended theory of gravity as in general relativity with the new or effective pressure and energy density as in \eqref{eq:14}-\eqref{eq:15}. We can calculate the energy conditions such as: NEC (null energy condition, WEC (weak energy condition), SEC (strong energy condition) and DEC (dominant energy condition) for the GRH bouncing model as,
\begin{eqnarray}
\rho+p &=& \frac{\eta^{2} \left[-3\Gamma \epsilon(1+2\beta)\eta^{2}t^{2}+(1-\eta^{2}t^{2})[2(1+3\beta)(1+\Gamma \epsilon)-3\beta]\right]}{(1+\Gamma \epsilon)(1+8\beta^{2}+6\beta)(1+\eta^{2}t^{2})^{2}} \geq 0 \nonumber\\
\rho+p &=&\frac{\eta^{2} \left[-3\Gamma \epsilon(1+2\beta)\eta^{2}t^{2}+(1-\eta^{2}t^{2})[2(1+3\beta)(1+\Gamma \epsilon)-3\beta]\right]}{(1+\Gamma \epsilon)(1+8\beta^{2}+6\beta)(1+\eta^{2}t^{2})^{2}} \geq 0; \rho \geq 0 \nonumber\\
\rho+3p&=&\frac{\eta^{2}\left[3(2+\Gamma \epsilon)(1+2\beta)\eta^{2}t^{2}-(1-\eta^{2}t^{2})[2(1+3\beta)(1+\Gamma \epsilon)+3\beta]\right]}{(1+\Gamma \epsilon)(1+8\beta^{2}+6\beta)(1+\eta^{2}t^{2})^{2}} \geq0 \nonumber \\
\rho-p&=&\frac{n^{2}\left[3(1+2\beta)\eta^{2}t^{2}(1-3(1+\Gamma \epsilon))+3(1-\eta^{2}t^{2})[2(1+3\beta)(1+\Gamma \epsilon)-3\beta]\right]}{(1+\Gamma \epsilon)(1+8\beta^{2}+6\beta)(1+\eta^{2}t^{2})^{2}}\nonumber \\
&\geq&0 \label{eq:17}
\end{eqnarray}
Graphically, we have represented the behaviour of energy conditions in Fig. \ref{Fig.4}. The SEC and the NEC both are violating at the bounce. In the flat FRW universe, at the bouncing point $\rho \geq 0,$   $\rho+p < 0$. The energy density of the fluid decreases with contraction. It is constant at the bouncing point and then increases with subsequent expansion. Therefore, at the bounce, NEC is violated and the universe is accelerating after the bounce due to $\rho+3p <0$.

\section{Discussions and Conclusion}\label{SectionV}
The simple candidate for the dark energy model is the cosmological constant. The cosmological observations are providing more accurate information on the dark energy properties. So, theoretically further validate the cosmological model, the geometrical diagnostic for dark energy in the form of state finder pair would be more appropriate\cite{Sahni03}. It consists of two parameters the jerk parameter $j$ and snap parameter $s$ which can be obtained by performing the third and fourth order derivative respectively of scale factor. Both these quantities can be derived as,
 
\begin{eqnarray}
j&=&\frac{\dddot{a}}{aH^2}= -\frac{3}{t(1+\eta^{2}t^{2})} \label{eq:18}\\
s&=&\frac{j-1}{3(q-\frac{1}{2})}=
\frac{2\eta^2t(3+t+\eta^2t^3)}{3(2+3\eta^2t^2+\eta^4t^4}\label{eq:19}
\end{eqnarray}
	
\begin{figure}[h!]
\includegraphics[width=60mm]{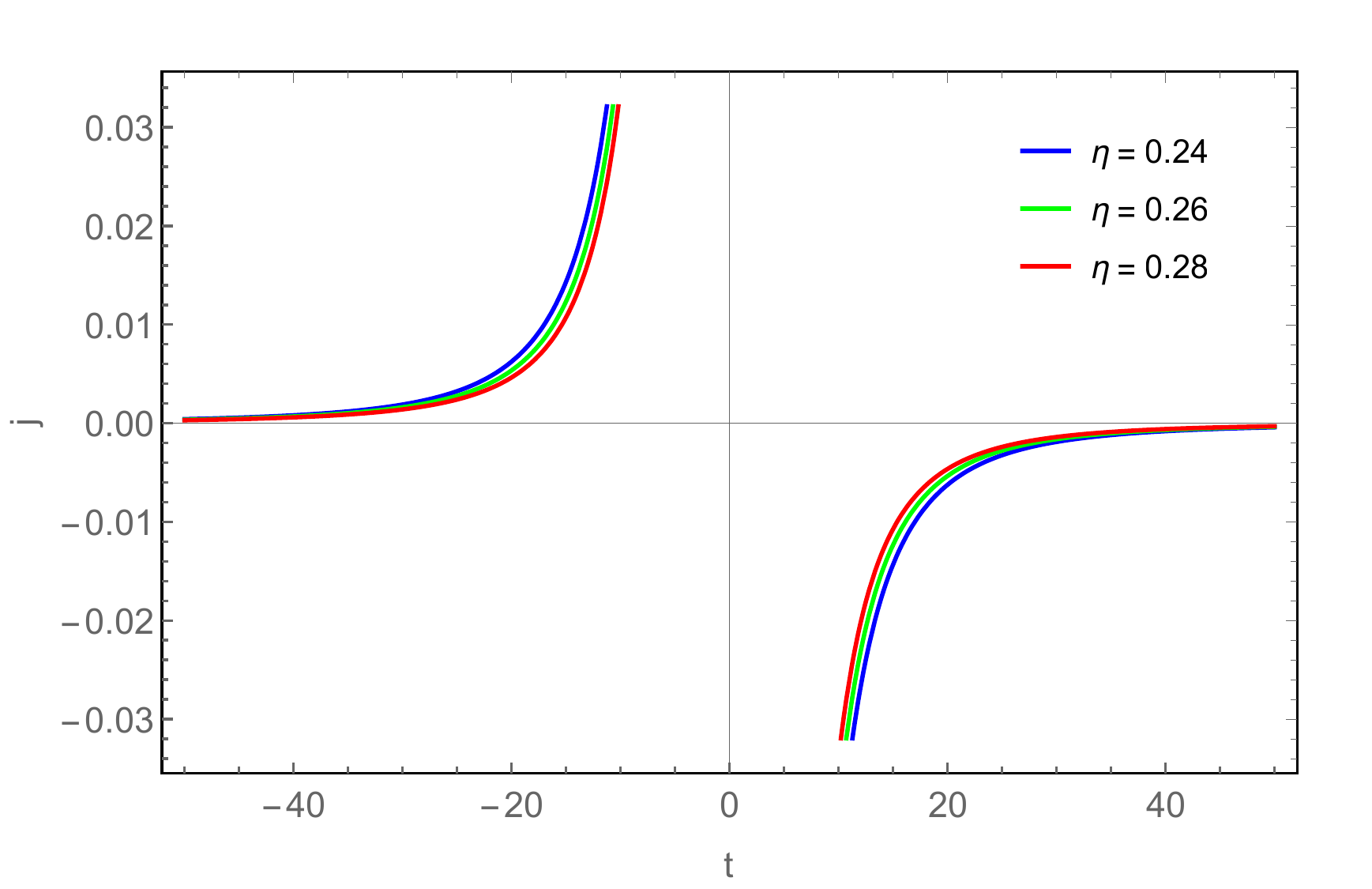}
\includegraphics[width=60mm]{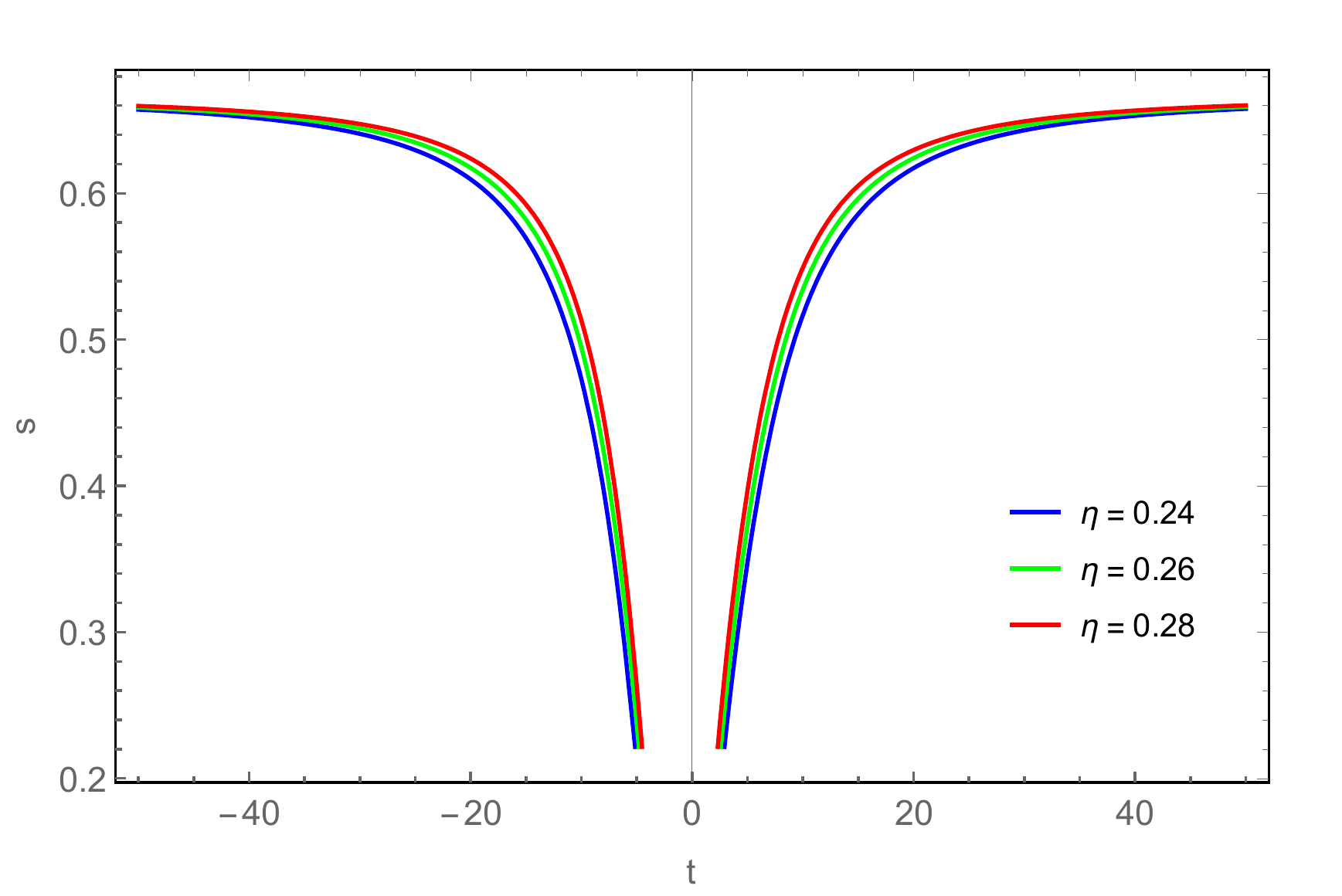}
\caption{Graphical representation of jerk parameter (left panel) and snap parameter (right panel) in cosmic time for the representative values of $\eta$.}\label{Fig.5}
\end{figure}
From eqn. \eqref{eq:19}, We can see that for $j=1$, the snap parameter should vanish and when the pair $(j,s)$ approaches to $(1,0)$, the model supports the $\Lambda$CDM behaviour. But from eqn. \eqref{eq:18}, it is evident that $j$ will approach to $1$ only in the negative time zone. Therefore, the model does not support the $\Lambda$CDM behaviour. The same behaviour can be observed from Fig. \ref{Fig.5} graphically. In the left panel, the jerk parameter is vanishing only at very early and very late time. Both the jerk and snap parameters are showing the singularity behaviour at the bouncing epoch. Finally, while framing this cosmological model assumptions were made to solve the system because of the difficulty in the governing equations. The physical viability of these assumptions must be known. In this particular problem, we have studied the model with an assumed scale factor. Therefore it is essential to know the stability of the model. The stability of the model can be studied through the mechanical stability of the cosmic fluid. This can be performed by the adiabatic speed of sound through the cosmic fluid as, $C_s^2=\frac{dp}{d\rho}$ \cite{Balbi07,Xu12,Mishra21}, which can be calculated for this model as,

\begin{equation}\label{eq:20}
C^{2}_{s}=(1+\Gamma \epsilon)\left[\frac{-3(1+2\beta)(1+\eta^{2}t^{2})^{2}-8\alpha \eta^{2}t}{3(1+2\beta)(1+\eta^{2}t^{2})^{2}+8\beta \eta^{2}t}\right] 
\end{equation} 	
				
\begin{figure}[h!]
\centering
\includegraphics[width=90mm]{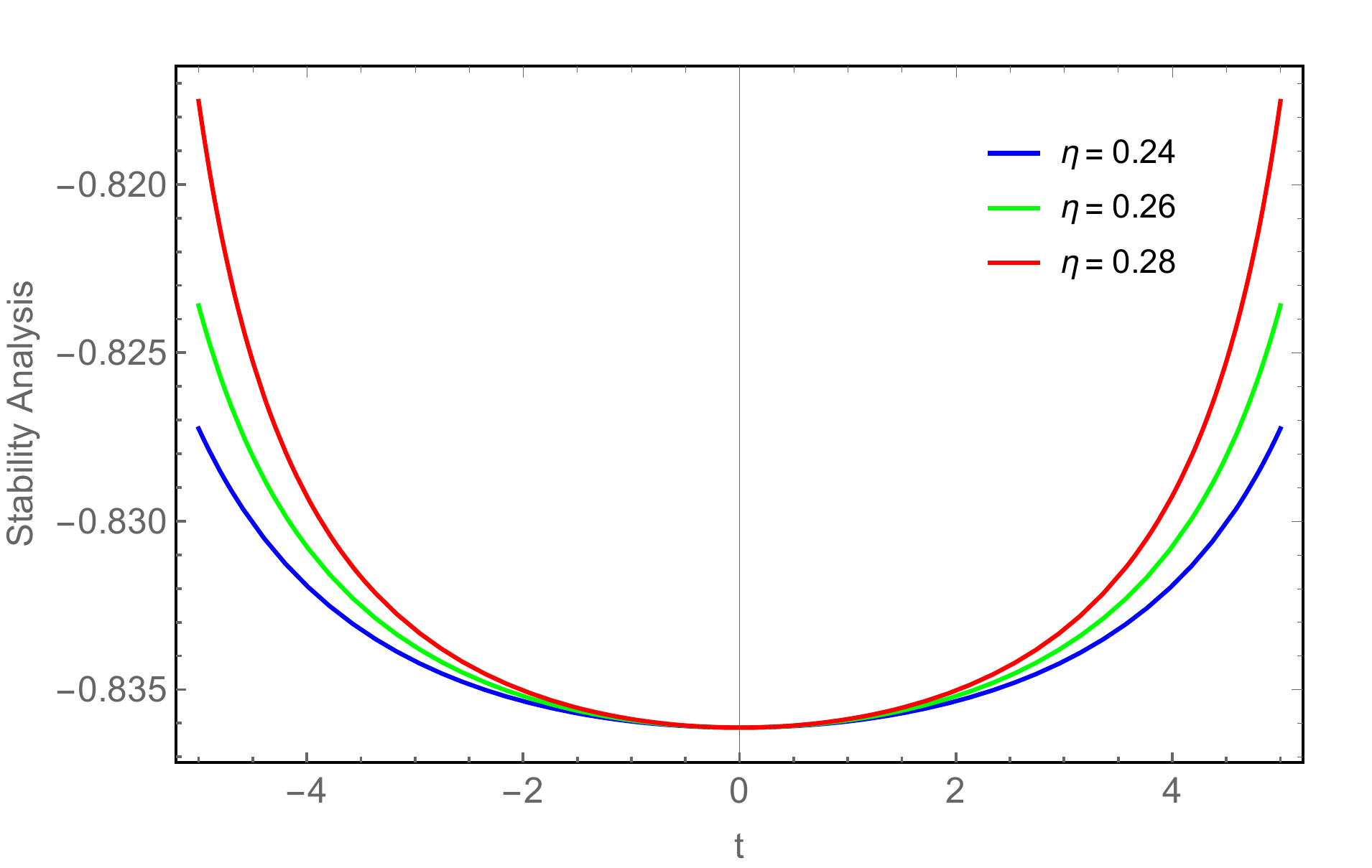}
\caption{Plot of stability analysis versus cosmic time for the representative values of $\eta$, $\beta=-0.51,$ $\Gamma=2,$ $\epsilon=0.1$} \label{Fig.6}
\end{figure}
The stability of the model depends on the behaviour of $C_s^2$ with respect to the cosmic time. If $C_s^2>0$ the model is stable and for $C_s^2<0$, the model remains unstable. In this model, we wish to know the stability at the bouncing point. Fig. \ref{Fig.6}  stability factor shows symmetry behaviour and at the bouncing point and in its neighbourhood , it remains negative. Hence we can conclude that the GRH bouncing model presented here is unstable.\\

The bouncing cosmological model constructed here in presence of GRH in an extended theory of gravity provides non-singular bounce at $t=0$.  The energy density increases and remains maximum at the bounce epoch and then decreases. Higher the value of $\eta$, the energy density is more. The violation of NEC at the bounce further gurantees the violation of SEC. The EoS parameter remains at the quintessence phase at the bounce and showing the well-shaped, but while evolving out, it gradually increases both in the negative and positive time zone. From the geometrical diagnostic, the models does not support $\Lambda$CDM behaviour, which has already been observed from the EoS parameter. The behaviour of the deceleration parameter shows the super-exponential expansion. Some of the key results of the model has been listed down in TABLE-I

\begin{table}[htb] 
\centering 
\begin{tabular}{c c c c c} 
\hline\hline 
\textbf{Parameter} &\quad \quad \textbf{$(t<<0)$)} & \quad \quad \textbf{($t=0$)}  & \quad \quad \textbf{ $(t>>0)$} & \\ [0.5ex] 
\hline
Scale Factor &\quad \quad Positive, Decreasing &  \quad \quad Positive, Fixed & \quad \quad Positive, Increasing & \\
Hubble Parameter & \quad \quad Negative, Increasing &\quad \quad Vanishes & \quad \quad Positive, Increasing & \\
Deceleration Parameter &\quad \quad Negative, Decreasing &\quad \quad Singularity Occurs & \quad \quad Negative, Increasing & \\
EoS Parameter& \quad \quad Negative, Decreasing &\quad \quad Negative, Fixed & \quad \quad Negative, Increasing &\\
Jerk Parameter& \quad \quad Positive, Increasing &\quad \quad Singularity Occurs & \quad \quad Negative, Increasing &\\
Snap Parameter & \quad \quad Negative, Decreasing &\quad \quad Singularity Occurs & \quad \quad Negative, Increasing & \\
\hline
\hline
\textbf{Energy Conditions} & \quad \quad\textbf{$(t<<0)$)} & \quad \quad \textbf{($t=0$)}  & \quad \quad \textbf{ $(t>>0)$} &\\
\hline
NEC,WEC& \quad \quad Satisfied & \quad \quad Satisfied & \quad \quad Violated & \\
DEC& \quad \quad Satisfied & \quad \quad Satisfied & \quad \quad Satisfied &\\
SEC& \quad \quad Violated & \quad \quad Violated &  \quad \quad Violated & \\
\hline 
\end{tabular}
\caption{The Key results of the model and the behaviour of cosmological parameters }
\label{table:nonlin} 
\end{table}
In conclusion, we say that the model is not stable at the bounce, which is in line with the statement that bouncing models suffer with instability. The other basic requirements of the bouncing model in an extended theory of gravity are satisfied.

\section*{Acknowledgements}
ST acknowledges Rashtriya Uchachatar Shikshya Abhiyan(RUSA), Ministry of HRD, Govt. of India for the financial support. The authors are thankful to the honorable reviewers for their valuable suggestions and comments for the improvment of the manuscript.

\end{document}